\documentclass[aip,jcp,amsmath,amssymb,reprint]{revtex4-1}
\usepackage[utf8]{inputenc} 
\usepackage[T1]{fontenc}       
\usepackage{amssymb}
\usepackage{amsmath}
\usepackage{comment} 

\usepackage{graphicx} 
\usepackage{color}

\usepackage{booktabs} 
\usepackage{array} 
\usepackage{paralist} 
\usepackage{verbatim} 
\usepackage{subfig} 
\usepackage{hyperref}
\usepackage[version=3]{mhchem}
\usepackage{feynmp}
  
\makeatletter
\newcommand*{\rom}[1]{\expandafter\@slowromancap\romannumeral #1@}
\makeatother

\begin{document}
\title{Time dependent second order Green's function theory for neutral excitations} 

\author{Wenjie Dou}
\email{douwenjie@westlake.edu.cn}
\affiliation{School of Science, Westlake University, Hangzhou, Zhejiang 310024, China}
\affiliation{Institute of Natural Sciences, Westlake Institute for Advanced Study, Hangzhou, Zhejiang 310024, China}

\author{Joonho Lee}
\email{jl5653@columbia.edu}
\affiliation{Department of Chemistry, Columbia University, New York, New York 10027, USA}

\author{David R. Reichman}
\email{drr2103@columbia.edu}
\affiliation{Department of Chemistry, Columbia University, New York, New York 10027, USA}

\author{Roi Baer}
\email{roi.baer@huji.ac.il}
\affiliation{Fritz Haber Center for Molecular Dynamics, Institute of Chemistry, The Hebrew University of Jerusalem, Jerusalem 91904, Israel}

\author{Eran Rabani}
\email{eran.rabani@berkeley.edu}
\affiliation{Department of Chemistry, University of California, Berkeley, California 94720, USA}
\affiliation{Materials Sciences Division, Lawrence Berkeley National Laboratory, Berkeley, California 94720, USA}
\affiliation{The Raymond and Beverly Sackler Center of Computational Molecular and Materials Science, Tel Aviv University, Tel Aviv 69978, Israel}

\begin{abstract}
We develop a time dependent second order Green's function theory (GF2) for calculating neutral excited states in molecules. The equation of motion for the lesser Green's function (GF) is derived within the adiabatic approximation to the Kadanoff-Baym (KB) equation using the second order Born approximation for the self-energy. In the linear response regime, we recast the time dependent KB equation into a Bethe-Salpeter-like equation (GF2-BSE), with a kernel approximated by the second order Coulomb self-energy. We then apply our GF2-BSE  to a set of molecules and atoms and find that GF2-BSE is superior to configuration interaction with singles (CIS) and/or time dependent Hartree-Fock (TDHF), particularly for charge transfer excitations, and is comparable to CIS with perturbative doubles (CIS(D)) in most cases. 
\end{abstract}

\maketitle
\section{Introduction}
Calculating excited states in molecules remains one of the grand challenges for computational chemistry. Time dependent Hartree-Fock (HF) \cite{mclachlan1964time,hirata1999configuration} and Time dependent Density Functional theory (DFT) \cite{Runge1984,hirata1999time,Leeuwen2001,Onida2002,Maitra2002,Marques2004,Jacquemin2009,Casida2009} offer affordable and simple means to calculate excited states energies, but suffer from accuracy with the corresponding excitation energies deviating by up to several eVs. Other quantum chemistry methods, such as coupled cluster (CC) with singles and doubles within the equation-of-motion formalism (EOM-CCSD), are more accurate, but are limited to relatively small system sizes due to steep computational scaling. 

An alternative to the above is based on many-body perturbation theory (MBPT) within Green's function theory. In particular, the so called ``GW'' approximation and the Bethe-Salpeter equation (BSE) have been successful in predicting charge and optical excitation, respectively,  particularly for solids. \cite{Albrecht1998,Benedict1998,Rohlfing2000,Benedict2003,Spataru2004,Tiago2006,sai2008optical,rocca2012solution,Fuchs2008,Ramos2008,Palummo2009,Schimka2010,Rocca2010,Blase2011,Faber2012,faber2014excited} However, the application of MBPT is often limited to simple solids with small unit cells due to the steep computational costs. This has led to many efforts in recent years aiming to reduce the computational cost of MBPT-based techniques. \cite{PhysRevB.75.235102,PhysRevB.88.075105,PhysRevLett.113.076402,rabani2015time,PhysRevB.85.081101,deslippe2012berkeleygw,foerster2011n,gonze2009abinit} 

The second order Green's function (GF2) approach falls into this latter category. It has, however, remained somewhat less popular than the GW/BSE approaches. {Recently, GF2 theory has experienced a renaissance \cite{phillips2014communication,pavovsevic2017communication,ohnishi2016explicitly}, partially due to its simplicity and the inclusion of dynamical exchange effects. } In GF2, the self-energy is described by the second order Born approximation,\cite{stefanucci2013nonequilibrium,kananenka2016efficient} resulting in a class of dynamical exchange effects~\cite{cederbaum1975one,Holleboom-1990,dahlen2005variational} that appear only at second and higher orders, and thus are often ignored in GW/BSE. On the other hand, GF2 treats the polarization term differently than GW/BSE and its accuracy in describing excited states in molecules is unexplored. The inclusion of such dynamical exchange terms in GF2 leads to $O(N^5)$ scaling of the computational cost where $N$ is size of the system.  

To overcome this computational bottleneck of GF2, we have recently introduced a stochastic approach to GF2 (sGF2), based on a stochastic resolution of identity used to decouple the $4$-index Coulomb integrals.\cite{Takeshita-2017stochastic,takeshita2019stochastic,dou2019stochastic} Unlike the standard resolution of identity (RI),\cite{Whitten-1973,Dunlap-1979-2,Dunlap-1979,Vahtras-1993,Feyereisen-1993} the number of stochastic orbitals (playing the role of the resolution basis) does not increase with the system size for certain ground state properties~\cite{takeshita2019stochastic} and for charge excitations.\cite{dou2019stochastic} This allows one to reduce the computational scaling of sGF2 theory to $O(N^3)$ at the expense of introducing a controlled statistical error in the calculated observable. Similar stochastic approaches to electronic structure theory have been developed also for other frameworks, including GW,\cite{PhysRevLett.113.076402,vlvcek2018swift} DFT,\cite{Rabani-2013-3,Rabani-2014-4,cytter2014metropolis,chen2019overlapped} MP2,\cite{neuhauser2012expeditious,ge2013guided,Takeshita-2017stochastic,Schafer2018Feb} RPA,\cite{Rabani-2013-2} and  more recently have also been used to improve the performance of auxiliary-field quantum Monte Carlo.\cite{lee2020stochastic}

In previous work we have used the imaginary time and real time sGF2 theory to calculate the ground-state and quasi-particle properties of molecular systems and assessed the accuracy of the stochastic resolution of identity for GF2.\cite{takeshita2019stochastic,dou2019stochastic} Furthermore, we have introduced a range-separated stochastic RI approach, where the long-range Coulomb term was approximated by the stochastic RI and the short-range Coulomb term was described by a deterministic RI.\cite{dou2020range} This hybrid deterministic-stochastic range-separated RI reduces the stochastic noise while maintaining the $O(N^3)$ computational scaling, which allows us to calculate the quasi-particle spectrum for systems containing $N=1000$ electrons or more.\cite{dou2020range}   

In the present work, we extend the GF2 formalism to describe neutral excitations by considering explicitly the optical field within a time dependent nonequilibrium GF2 theory. Unlike the equilibrium case, non-equilibrium Green's functions depend on two times and thus are more challenging to compute. Here, to simplify the calculation, we first invoke an adiabatic approximation to obtain the self-energy. We then derive within linear response theory a Bethe-Salpeter-like equation along with the second order Born approximation to describe the Bethe-Salpeter kernel. {\color{black} While our long term goal is to go beyond the adiabatic approximation and to develop a stochastic approach to reduce the scaling of the GF2 theory for excited states to $O(N^3)$, here we focus on assessing the accuracy of GF2 for neutral excitations and comparing the approach to other methods such as second-order approximation to EOM-CCSD (EOM-CC2) and configuration interaction with singles and perturbative doubles (CIS(D)). We note in passing that while it is difficult to formulate a stochastic version of EOM-CCSD with favorable scaling (as compared to the $O(N^3)$ of a stochastic GF2), two of us are formulating a stochastic version of EOM-CC2, which will be comparable to GF2 in terms of scaling.} 

We show that at least in some simple but paradigmatic examples, that the time dependent GF2 theory provides a more accurate framework to predict excited states compared to TDHF and configuration interaction with singles (CIS), and is competitive with and CIS(D) and EOM-CC2. In particular, for charge transfer states, CIS fails to predict the correct energy, whereas sGF2 provides a more reliable estimate of charge transfer excitations. 

The manuscript is organized as follows: In Sec.~\ref{sec:theory} and Sec.~\ref{sec:weak}, we present the formulation of the time dependent GF2 theory in both the time and frequency domains. In Sec.~\ref{sec:results} we test the performance of GF2 theory against other quantum chemistry methodologies for a representative set of atoms and molecules. Finally, in Sec.~\ref{sec:con} we conclude. 

\section{Theory} \label{sec:theory}
\subsection{Notation}
We start by defining a general electronic Hamiltonian in second quantization form. The formulation is general and applies to any choice of basis set. In this section we us the notation $i, j, k, l\cdots$ to represent the indices in a general basis. The Hamiltonian is given by
\begin{equation} \label{eq:Hami}
\hat H = {\hat H}_0 + \sum_{ij} \Delta_{ij}(t) \hat a^\dagger_i \hat a_j,
\end{equation} 
where  $\hat a^\dagger_i$ ($\hat a_i$) is the creation (annihilation) operator for an electron in orbital $\chi_i(\bf r)$, 
\begin{equation} \label{eq:Hami0}
{\hat H}_0 = \sum_{ij} h_{ij}   \hat a^\dagger_i \hat a_j + \frac12 \sum_{ijkl}v_{ijkl}  \hat a^\dagger_i \hat a^\dagger_k  \hat a_l \hat a_j 
\end{equation} 
is the unperturbed Hamiltonian, $h_{ij}$ is the one-body matrix element in the basis, and $v_{ijkl}$ represents the two-body 4-index Coulomb integral, given by
\begin{equation} 
	\label{eq:2e4c}
	v_{ijkl} = (i  j | kl ) = \iint  \frac{\chi_i ({\bf r}_1)\chi_j ({\bf r}_1)\chi_k ({\bf r}_2)\chi_l ({\bf r}_2)}{\left|{\bf r}_1-{\bf r}_2\right|} d{\bf r}_1 d{\bf r}_2.
\end{equation}
The last term  on the right hand side of Eq.~\eqref{eq:Hami} represents a time dependent perturbation, which is assumed to be one-body, suitable for describing the linear absorption spectrum, where $\Delta_{ij}(t)$ is the time dependent one-body matrix element of the perturbation. In situations where the chosen basis is not orthonormal, we also define the overlap $S$ with the overlap matrix elements of orbital $\chi_i ({\bf r})$ and $\chi_j({\bf r})$ as
\begin{eqnarray}
 S_{ij} = (i|j) = \int \chi_i ({\bf r})\chi_j ({\bf r}) d{\bf r}.
\end{eqnarray}

\subsection{Green's Function and Kadanoff-Baym Equations}
The traditional approach to describe excited states generated by the Hamiltonian~\eqref{eq:Hami} is based on solving the many-body eigenvalue problem, $H |\Psi_n \rangle = E_n | \Psi_n \rangle$, where $E_n$ and $|\Psi_n\rangle$ are the n$th$ eigenvalue and eigenstate, respectively. A complete solution within full configuration-interaction (FCI) is prohibitive for large system sizes or large atomic basis sets, and thus, most excited state calculations are based on introducing approximations with reduced computational scaling. 

The Green's function formalism offers a systematic way to treat the many-body interactions in Eq.~\eqref{eq:Hami} using diagrammatic expansions. A central quantity in this approach is the single particle lesser Green's function defined as (we assume that $\hbar=1$)
\begin{eqnarray}
G_{ij}^< ({t}_1, {t}_2)  = i \langle \hat a^\dagger_j (t_2) \hat a_i
({t}_1)  \rangle.
\label{eq:Gij}
\end{eqnarray}
In the above, we have used the Heisenberg representation for the time dependent operator 
\begin{eqnarray}
\hat a^\dagger_j ({t}) = {\cal T} e^{i \int_0^t \hat H ({t'}) dt' } \hat a^\dagger_j e^{-i \int_0^t \hat H({t'}) dt'},
\end{eqnarray}
where ${\cal T}$ is the time-ordering operator and the expectation value is calculated within the grand-canonical ensemble: $\langle \cdots \rangle = Z^{-1} \mbox{Tr} \left[ (\cdots)e^{-\beta ({\hat H}_0 - \mu \hat N) } \right] $. Here, $Z=\mbox{Tr} \left[ e^{-\beta ({\hat H}_0 - \mu \hat N)} \right]$ is the grand-canonical partition function, $\beta$ is the inverse temperature, $\mu$ is the chemical potentials, and $\hat N=\sum_{i} \hat a^\dagger_i \hat a_i$ is the number operator. 

The equations of motion for the lesser Green's function follow the Kadanoff-Baym equations:\cite{stefanucci2013nonequilibrium} 
\begin{eqnarray}
i S \partial_{t_1} G^<(t_1, t_2) =  F[\rho(t_1)] G^< (t_1, t_2 ) + I^< (t_1, t_2), \label{eq:KB1}
\end{eqnarray}
and
\begin{eqnarray}
-i  \partial_{t_2} G^<(t_1, t_2) S =  G^< (t_1, t_2 ) F[\rho(t_2)] - I^< (t_2, t_1)^*, \label{eq:KB2}
\end{eqnarray}
where $\rho(t)=-i G^<(t, t)$ is the density matrix and $F[\rho]$ is the Fock operator with matrix elements
\begin{eqnarray}
\label{eq:HF+Delta}
F_{ij}[\rho] &=& h_{ij} + v^H_{ij}[\rho] + v_{ij}^{x} [\rho] + \Delta_{ij}(t).
\end{eqnarray}
In the above, $v^H_{ij} [\rho] = \sum_{kl} v_{ijkl} \rho_{kl}$ is the matrix element of the Hartree potential and $v^{x}_{ij} = \sum_{kl} v_{ikjl} \rho_{kl}$ is the matrix element of the exchange interaction. Finally, the scattering integral, $I^< (t_1, t_2)$, appearing in Eqs.~\eqref{eq:KB1} and \eqref{eq:KB2} is given by:
\begin{equation}
\label{eq:I<}
\begin{split}
I^< (t_1, t_2) &= \int_{0}^{t_1} \Sigma^R(t_1, t_3) G^< (t_3, t_2)dt_3 + \\
&\int_{0}^{t_2} \Sigma^<(t_1, t_3) G^A (t_3, t_2)  dt_3. 
\end{split}
\end{equation}
In the above equation, $G^A(t_1,t_2)=i \theta(t_2-t_1) \langle \hat a^\dagger_j (t_2) \hat a_i ({t}_1)  \rangle$ is the advanced Green's function and $\theta(t)$ is the Heaviside step-function. The exact form of the retarded ( $\Sigma^R(t_1, t_2)$) and lesser ($\Sigma^<(t_1, t_2)$)  self-energies is difficult to obtain apart from the case of  simple model systems and thus most GF calculations are based on approximating the self-energies using different closures. In this work we resort to the second order Born approximation to compute the self-energies and, as will become clear below, we describe only the retarded self-energy appearing above (Langreth rules have been used~\cite{stefanucci2013nonequilibrium})
\begin{eqnarray} \label{eq:sigmaR}
\Sigma^R_{ij}(t_1, t_2 ) &= \sum_{kl} i G^<_{kl}(t_1, t_2 ) \delta W^R_{ikjl}(t_1, t_2 )\nonumber \\ 
&+i G^R_{kl}(t_1, t_2 )\delta W^>_{ikjl}(t_1, t_2 ),
\end{eqnarray}
where the retarded and greater screened Coulomb integrals are given by
\begin{widetext}
 \begin{eqnarray} \label{eq:WR}
\delta W^R_{ikjl}(t_1, t_2 ) &=& -i  \sum_{mnqp}   ( G^<_{mn}(t_1, t_2 ) G^A_{qp}(t_2, t_1 )  
+ G^R_{mn}(t_1, t_2 ) G^<_{qp}(t_2, t_1 )  )v_{impk} (2 v_{jnql} - v_{jlqn} ),   \\
\delta W^>_{ikjl}(t_1, t_2 ) &=& -i  \sum_{mnpq}  G^>_{mn}(t_1, t_2 ) G^<_{qp}(t_2, t_1 ) v_{impk} (2 v_{jnql} - v_{jlqn} ),  
\end{eqnarray}
\end{widetext}
and, as before, $G^{R,A}_{ij}(t_1,t_2)$ and $G^{<,>}_{ij}(t_1,t_2)$ are the retarded/advanced and the lesser/greater Green's function, respectively.

\subsection{The adiabatic approximation}
Before we proceed, we would like to point out to another challenge associated with the need to propagate the GF and to evaluate the self-energies along two times. In the following subsection, we will invoke the adiabatic approximation similarly to the approach taken for GW/BSE.\cite{attaccalite2011real}   This allows us to reduce the complexity associated with describing  two-time self-energies. The idea behind the adiabatic approximation is that the system responds instantaneously to the external driving force such that the integral in Eq.~\eqref{eq:I<} becomes local in time.

It is convenient to define the central time $t$ and the time difference $\tau$ 
\begin{eqnarray}
\label{eq:t,tau}
t &=& (t_1 + t_2 )/2,\\
\tau &=& t_1 - t_2,
\end{eqnarray}
and express the self-energies appearing in the scattering integral (cf., Eq.~\eqref{eq:I<}) by \cite{attaccalite2011real} 
\begin{eqnarray} \label{eq:adiabatic}
\Sigma^R (t_1, t_2)  &\approx&\tilde \Sigma^{\rm ad} (t) \delta(\tau), \\
\Sigma^<  (t_1, t_2) &\approx& 0 .
\end{eqnarray}
In the above equation, $\tilde \Sigma^{\rm ad} (t)$ is defined as
\begin{eqnarray}
\label{eq:sigmaad}
 \tilde \Sigma^{\rm ad} (t) &= \int \Sigma^R (t_1, t_2) e^{i \omega \tau} d\tau \equiv  \tilde \Sigma^R (t, \omega).
\end{eqnarray}
{\color{black}The adiabatic approximation is consistent with taking the $\omega\rightarrow 0$ limit assuming that the plasma energy (the main source of screening) is much higher than the neutral excitation energy differences.\cite{PhysRevB.69.205204}}
Note that in the above equation, $\tilde \Sigma^{\rm ad} (t)$ is given in terms of a Fourier transform of the time difference variable, $\tau$ denoted by ``tilde'' ($\tilde{\cdots}$). 

Using the adiabatic approximation for the self-energy, the equation of motion for the equal time  Green's function, $-iG^< (t_1=t, t_2=t) = \rho(t)$, can be simplified as
\begin{eqnarray}
i \partial_{t} \rho(t)  &= S^{-1}( F[\rho(t)] +  \tilde{\Sigma}^{\rm ad}(t) ) \rho(t) \nonumber \\ &- \rho(t)( F[\rho(t)]  + \tilde{\Sigma}^{\rm ad}(t) )S^{-1}, {~}{~}{~}{~}
\label{eq:rho}
\end{eqnarray}
where $\rho(t)$ is the density matrix and $\tilde{\Sigma}^{\rm ad}(t)$ is defined in Eq.~\eqref{eq:sigmaad}. In principle, the solution of the above equation requires a knowledge of the retarded and lesser Green's functions to obtain $\tilde{\Sigma}^{\rm ad}(t)$ from Eq.~\eqref{eq:sigmaad}. While this seems challenging, the adiabatic limit offers a significant simplification within the weak perturbation limit, as described in the following section.  

\section{The Weak driving limit} \label{sec:weak}
In the limit of weak external perturbation, namely, when $\Delta_{ij}(t) \rightarrow 0$, we can further simplify the description of the self-energy and recast the time dependent
equation of motion for the density matrix into a Casida-like form. In this section, we first provide a working expression to obtain the  $\tilde{\Sigma}^{\rm ad}(t)$ without solving the Kadanoff-Baym equations for the retarded and lesser Green's functions, and then derive a Casida-like equation to describe the neutral (excitonic) spectrum within the GF2 closure. 
\subsection{Self-energy}
\label{subsec:sigma-weak}

So far, the formalism described above makes no assumption about the basis set used. However, using the  eigenvalues of Eq.~\eqref{eq:HF+Delta} with $\Delta_{ij}(t)=0$ leads to a significant simplification. Thus, we now describe the calculation of $\Sigma^{\rm ad}(t)$ using the eigenstates ($\psi_i$) and the eigenvalues ($\varepsilon_i$) of the Fock matrix. To obtain an expression for $\Sigma^{\rm ad}(t)$ we treat the two terms appearing on the right hand side of Eq.~\eqref{eq:sigmaR} separately, and refer to the corresponding self-energies with superscripts ``ad1'' and ``ad2''. For $\tilde \Sigma^{\rm ad1}_{ij}(t)$ we further use an approximation~\cite{attaccalite2011real,PhysRevB.69.205204} for the lesser Green's function given by $G_{ij}^<(t,\omega)  = i 2\pi \delta( \omega - \varepsilon_i) G^<_{ij}(t, \tau=0) $. This leads to the following expression for $\tilde \Sigma^{\rm ad1}_{ij}( t)$ (see Appendix \ref{app:a} for more details)
\begin{eqnarray}
\tilde \Sigma^{\rm ad1}_{ij}( t) &\approx i \sum_{kl} \delta \tilde W^R_{ikjl}(- \varepsilon_k) G^<_{kl} (t, \tau=0) \nonumber \\ &=- \sum_{kl} \delta \tilde W^R_{ikjl}(- \varepsilon_k) \rho_{kl} (t),
 \end{eqnarray}
where, as before, $\rho_{kl} (t)$ is the $kl$ matrix element of the density matrix. Following similar steps for $\tilde \Sigma^{\rm ad2}_{ij}(t)$ we obtain (see Appendix \ref{app:a} for more details)
\begin{eqnarray} 
\tilde \Sigma^{\rm ad2}_{ij}  (t) \approx  \frac12  \Re \sum_{kl} \delta \tilde W^R_{ikjl} (0) \delta_{kl}.  
\end{eqnarray} 
Note that $\tilde \Sigma^{\rm ad2}_{ij}  (t)$ is time-independent. Using the zeroth order approximation to the Green's functions appearing in Eq.~\eqref{eq:WR}, we find that
\begin{widetext}
 \begin{eqnarray} \label{eq:W}
\delta \tilde W^R_{ikjl} (\omega) &&= \sum_{mnpq} [ f(\varepsilon_n) \frac{1}{\varepsilon_n  - \omega - \varepsilon_q - i\eta}  +  \frac{1}{\omega + \varepsilon_p - \varepsilon_n + i\eta}  f(\varepsilon_p)  ] \delta_{mn} \delta_{pq} v_{impk} (2 v_{jnql} - v_{jlqn} ) \nonumber \\
&&= \sum_{mnpq} \frac{f(\varepsilon_n) - f(\varepsilon_q) }{\varepsilon_n  - \omega - \varepsilon_q - i\eta}  \delta_{mn} \delta_{pq} v_{impk} (2 v_{jnql} - v_{jlqn} )
= \sum_{nq} \frac{f(\varepsilon_n) - f(\varepsilon_q) }{\varepsilon_n  - \omega - \varepsilon_q - i\eta}  v_{inqk} (2 v_{jnql} - v_{jlqn} ) .
\end{eqnarray} 
\end{widetext}
In the above equation $\eta$ is a small positive regularization parameter, and $f(\varepsilon_n)$ is the Fermi-Dirac distribution.


\subsection{Linear response theory: GF2-BSE}
For reasons that will become clear below, we denote the initial density matrix $\rho(t=0)\equiv\rho_0$ and the initial self-energy as $\Sigma^{\rm ad} (0) \equiv \Delta H$. Furthermore, 
\begin{widetext}
\begin{eqnarray} \label{eq:TD-GF2}
&&i \partial_{t} \rho
= S^{-1}\Big( F[\rho_0] + \Delta H+ v^H[\rho]  -  v^H[\rho_0] + v^{x}[\rho] - v^{x} [\rho_0] \Big) \rho \nonumber \\
&&- \rho \Big( F[\rho_0] + \Delta H+ v^H[\rho]  -  v^H[\rho_0] + v^{x}[\rho] - v^{x} [\rho_0] \Big) S^{-1} \nonumber \\
&&+ S^{-1} \Big (\Sigma^{\rm ad}(t)  -  \Sigma^{\rm ad}(t=0) \Big) \rho - \rho \Big (\Sigma^{\rm ad}(t)  -  \Sigma^{\rm ad}(t=0) \Big) S^{-1}, \nonumber \\
\end{eqnarray}
\end{widetext}
In the applications reported below, we obtain the quasiparticle-like correction term ($\Delta H$) from a stochastic GF2 calculation~\cite{dou2019stochastic} and the adiabatic self-energy ($\Sigma^{\rm ad}(t)$) using the second order Born approximation. The above equation can be used to describe both weak and strong driving forces, $\Delta(t)$ but can be further simplified into a Casida-like form within linear response theory (assuming weak driving forces). Before we describe the Casida-like form, in the following subsection we express $\Sigma^{\rm ad}(t)$ in terms of a screened Coulomb integral and the density matrix itself, using the second order Born approximation. This representation is essential to the applications within linear-response theory.

In the linear response regime, the time dependent GF2 equation can be recast into the usual Casida form.\cite{casida1995recent} Following the general guidelines used in time dependent DFT or TDHF,~\cite{Casida2009,hirata1999time,attaccalite2011real} the time dependent GF2 can be recast into a symplectic eigenvalue problem (assuming zero temperature and spin-restricted orbitals)
\begin{eqnarray} \label{eq:casida}
\begin{pmatrix}
A & B \\
-B & -A 
\end{pmatrix} 
\begin{pmatrix}
X  \\
Y 
\end{pmatrix} = \omega 
\begin{pmatrix}
X  \\
Y 
\end{pmatrix} ,
\end{eqnarray}
where $A$ and $B$ are matrices of size $N_{occ} N_{virt} \times N_{occ} N_{virt}$ and $N_{occ}$/$N_{virt}$ is the number of occupied/virtual molecular orbitals. These matrices can be expressed in terms of the exchange $K^X$ and direct $K^{DA}$ correlations
\begin{eqnarray}
A &=& D + 2 K^X  - K^{DA} \nonumber \\
B &=& 2 K^X - K^{DB}. 
\end{eqnarray}
The matrix elements of $D$, $K^X$, and $K^{DA}$ are given by
\begin{eqnarray}
D_{ia,jb} &&= (\omega_a - \omega_i) \delta_{ab} \delta_{ij},  \nonumber \\
K^X_{ia,jb} &&= \iint d{\bf r}_1 d{\bf r}_2 \frac{(\psi_i({\bf r}_1))^* 
\psi_a ({\bf r}_1)
(\psi_j({\bf r}_2))^* 
\psi_b ({\bf r}_2)}
{\left|{\bf r}_1-{\bf r}_2\right|}, \nonumber \\
K^{DA}_{ia,jb} &&= \delta \tilde W^R_{abij}  + v_{abij},  \nonumber \\
K^{DB}_{ia,jb} &&= \delta \tilde W^R_{ibaj}  + v_{ibaj}.  
\label{eq:DKxKd}
\end{eqnarray}
Here, $i,j$ represents occupied orbitals and $a, b$ are unoccupied orbitals. $\omega_a$ is the GF2 corrected energy in the molecular basis obtained by solving for\cite{dou2019stochastic}
\begin{eqnarray}
\label{eq:omega_n}
\omega_n = \varepsilon_n + \langle \psi_n | \Re \tilde \Sigma^R (\omega_n) | \psi_n \rangle,
\end{eqnarray}
where $\Sigma^R (\omega_n)$ is the self-energy obtain from a single time GF2 calculation.\cite{dou2019stochastic} Eqs.~\eqref{eq:casida}-\eqref{eq:DKxKd} are the main result of this subsection and are referred to as ``GF2-BSE''.  

\subsection{Computational scaling and outline of the approach}
The overall computational scaling of solving Eq.~\eqref{eq:TD-GF2} for $\rho(t)$ is determined by the computational scaling of the self-energy, which is the most expansive part of the calculation. Formally, the scaling of the self-energy is $O(N^5)$, where $N$ is the size of the basis. In linear response, the computational scaling for the GF2-BSE is $O(N^6)$, a common increase in scaling going from the time-domain to the frequency-domain.\cite{hirata1999time,attaccalite2011real}  A significant reduction in the overall scaling to cubic ($O(N^3)$) can be achieved using a stochastic resolution of identity~\cite{} applied to the time-domain formulation, which will be the subject of future study. Here, we simply will assess the accuracy of the deterministic GF2-BSE approach. Before doing so, we outline the main steps:
\begin{enumerate}
\item Perform a Hartree-Fock calculation and generate the Fock matrix (Eq.~\eqref{eq:HF+Delta}).
\item Use the single-time GF2 approach~\cite{dou2019stochastic} to calculate the self-energy appearing in Eq.~\eqref{eq:omega_n} and obtain the GF2 correction to the quasi-particle energies, $\omega_n$.
\item  Calculate $\delta \tilde W^R_{ik,jl} (\omega=0)$ using Eq.~\eqref{eq:W}. For the applications reported below we used a damping parameter ($\eta=0.01$ a.u.), which is sufficiently small to converge the results. 
\item Calculate the matrix elements of the Eq.~\eqref{eq:casida}, and solve for the eigenvalues and eigenstates. 
\end{enumerate}
Solving for $\omega_n$ is done self-consistently both for solving Eq.~\eqref{eq:omega_n} and also for obtaining $\tilde{\Sigma}^R(\omega)$. A simplification that often is useful within the framework of GW/BSE is to use a single-shot calculation, which amounts to replacing $\tilde{\Sigma}^R(\omega)$ with the zeroth order approximation.\cite{PhysRevB.69.205204} However, this may result in multiple solutions for $\omega_n$ and in  such cases, we restrict the solution to the fully self-consistent treatment.\cite{dou2019stochastic} We denote the former approach as G0F2-BSE and the later as GF2-BSE. 
\section{Applications to Atoms and Molecules} \label{sec:results}
In this section we compare results obtained using the GF2-BSE approach for the low lying excited states for a set of atoms and molecules, and compare GF2-BSE results to excited states obtained using CIS, TDHF, CIS(D), EOM-CCSD, and EOM-CC2 method. While EOM-CCSD is not an exact approach for excited states, its accuracy should be reasonable for low lying states.  In addition, we note that for two electron systems such as He ahd H$_{2}$, it is exact. We use the cc-pVDZ basis set for all cases discussed below and assume that the EOM-CCSD provides accurate references within this basis.

\begin{table} [h]
\caption{Difference of the lowest singlets (in eV) for CIS, TDHF, CIS(D), EOM-CC2, and G0F2-BSE compared to EOM-CCSD. }
\begin{tabular}{ c c c c c c c} 
 \hline \hline
    Atom         &   $\Delta$CIS      &$\Delta$TDHF      & $\Delta$CIS(D)   & $\Delta$EOM-CC2 & $\Delta$G0F2-BSE & \\ \hline 
He         &  -0.72     &-1.09       &-0.11 &-0.11   &-0.05  &\\ 
             & -0.82     &-0.98      &-0.15  &-0.15  & -0.67   &\\ 
             & -0.82     &-0.98       &-0.15  &-0.15     & -0.67   &\\ 
             & -0.82     &-0.98      & -0.15    & -0.15 & -0.67  &\\ \hline
             
Be        &  -0.34     &-0.64     &    -0.19 &-0.20 &-0.18&\\
            & -0.34     &-0.64     & -0.19  &-0.20 &-0.18 &\\
            & -0.34   &-0.64      &  -0.19  &-0.20 & -0.18&\\ \hline

Ne        & -1.05    & -1.19      &   0.14    &0.17 & 0.77\\ 
             & -1.05     & -1.19       & 0.14  &0.17 & 0.77\\ 
             & -1.05     & -1.19        & 0.14  &0.17  & 0.77\\ 
             & -1.09     & -1.16       &  0.14  &0.17 & 0.68 \\      
             & -1.09     & -1.16       &  0.14  &0.17& 0.68 \\   \hline
                               
Error  &   0.794 &	0.987	& 0.151	&0.167  & 0.522   \\ \hline\hline
\end{tabular} \label{table:2}
\end{table}

In Table~\ref{table:2}, we list several lowest singlet energy differences (compared to the corresponding EOM-CCSD singlet energies) for He, Be, and Ne. {\color{black} Note that all calculations performed are closed-shell calculations. The singlet energies may have degeneracy for these atoms, and we simply list them all}.  We find that the smallest deviations are observed for the CIS(D) approach, with an overall average error of $\approx 0.15$~eV. {\color{black} EOM-CC2 results are very similar to CIS(D) results, which is not too surprising given that both methods include electron correlation only up to the second order.} CIS and TDHF underestimate the excitation energies by an average of nearly $1$~eV while the G0F2-BSE approach outperforms CIS and TDHF with an average error of $0.5$~eV, however, G0F2-BSE a performs much better for the lowest singlet excitation. The G0F2-BSE underestimates the excitation energies in some cases and in others it overestimates them. The overall better performance of G0F2-BSE with respect to CIS/TDHF results from two main attributes: (a) the inclusion of screening effects in the G0F2-BSE kernel (missing from both CIS and TDHF) which reduces the coupling between electrons and holes and (b) the correction of the quasi-particle energies, $\omega_n$, that are assumed to equal the HF orbital energies in CIS/TDHF and thus overestimate the fundamental band gap (HOMO-LUMO gap). The error cancellation between the larger HF fundamental gap and the stronger Coulomb interactions give rise to overall reasonable excitation energies using CIS and TDHF. 

In Table~\ref{table:1}, we list the lowest singlet energy differences (compared to the corresponding EOM-CCSD singlet energies) for the vertical excitation of H$_2$ molecule within the cc-pVDZ basis. We find that for the two lowest singlet excitations, the G0F2-BSE provides slightly better agreement with EOM-CCSD compared to CIS, and TDHF, and only slightly under-performs in comparison to CIS(D) and EOM-CC2. The agreement between CIS, TDHF, EOM-CC2, CIS(D) and G0F2-BSE with the EOM-CCSD approach for higher excitations is rather poor. The higher excited states have a substantial ``double-excitation'' character which is described within the EOM-CCSD more accurately than the other methods (TDHF, CIS, and GF2-BSE do not account for double-excitations). 
\begin{table}[h] 
\caption{Difference of the several lowest vertical excitation energies (in eV) for H$_2$ using CIS, TDHF, CIS(D), EOM-CC2, and G0F2-BSE compared to the EOM-CCSD approach. The equilibrium bond distance is taken to be {0.74} \(\text{\AA}\)}
\begin{tabular}{ c c c c c c c} 
 \hline \hline
               $\Delta$CIS      &$\Delta$TDHF      &  $\Delta$CIS(D)  & $\Delta$EOM-CC2 & $\Delta$G0F2-BSE & \\ \hline 
                                
              0.15     &-0.01      & 0.11 &0.10 & 0.08 \\ 
              0.06    &-0.08       &  0.18  &0.16 & 0.28 \\ 
              2.92     &2.67         &  2.56 &2.59 & 2.48\\ 
              9.32    &9.18        & 9.63 & 9.62 & 9.34\\ 
              2.12     &1.98         & 2.43 &2.42 & 2.14\\ 
\hline\hline
\end{tabular} \label{table:1}
\end{table}

The results for atoms and the smallest molecule $\mbox{H}_2$ suggest that the GF2-BSE approach provides accurate excitation energy for the lowest excited state and somewhat less so for higher excitations. The systems studied so far support only one or two bound excited states at energy corresponding to the vertical transitions and the other excitations are often above the ionization threshold. This may explain the overall better performance of the GF2-BSE approach for the lowest excited state. Whether this is indeed the case can be tested for larger molecules, which may support many more bound excitations. 

In Tables~\ref{table:four} and \ref{table:3}, we list the vertical excitation energies for $5$ different molecules obtained using the aforementioned methods. To reduce the scaling and the computational effort, we use the stochastic GF2 methods~\cite{dou2019stochastic} to calculate the self-energies appearing in Eq.~\eqref{eq:omega_n} as well as to obtain the GF2 correction ($\Delta H$) to the quasi-particle energies, $\omega_n$. Thus, the vertical excitation energies obtained from the GF2-BSE method have an error bar resulting for the use of a stochastic approach. We have used $2000$ stochastic orbitals to obtain the full self-consistent self-energy. The stochastic errors from GF2-BSE are estimated using $10$ independent runs. {\color{black} Note that here we have only used the stochastic formulation of Matsubara and mixed time GF2 to obtain the self-energy and hence quasi-particle energies. A full stochastic implementation of the GF2-BSE theory will be presented and tested in future work. } 

\begin{table} 
\begin{center}
\begin{tabular}{ c c c c c c c c} 
 \hline \hline
             &   $\Delta$CIS      &$\Delta$TDHF      &  $\Delta$CIS(D)   &  $\Delta$EOM-CC2  & $\Delta$GF2-BSE & \\ \hline 

 LiH            
&0.57         &0.53         &0.24           &0.25      &0.03 $\pm$    0.04 \\
&(-0.13)         &(-0.09)         &(0.20)   &(0.19)              &(-0.08) \\
&0.60         &0.57         &0.25       &0.26            &0.09 $\pm$    0.04 \\
&0.60         &0.57         &0.25       &0.26            &0.11 $\pm$    0.04 \\
&0.45         &0.39         &0.26       &0.26          &0.15 $\pm$    0.03 \\
&0.49         &0.47         &0.24       &0.24          &-0.16 $\pm$    0.04 \\
&0.49         &0.47         &0.24       &0.24           &-0.13 $\pm$    0.04 \\
&0.43         &0.42        &0.19        &0.19        &-0.17 $\pm$    0.04 \\
&0.50        &0.40        &0.26         &0.27       &0.08 $\pm$   0.04 \\
&0.39        &0.34        &0.29         &0.28      &-0.36 $\pm$   0.04 \\
Error  &     0.500	& 0.462	& 0.248	 &0.249 & 0.142 &\\ \hline
 
HF             
&1.23        &1.16        &-0.20       & -0.14        &-0.33 $\pm$    0.16 \\
&(0.97)         &(1.04)         &(2.40) &   (2.54)             &(-0.16) \\
&1.23        &1.16        &-0.20    &  -0.14         &-0.17$\pm$    0.15 \\
&0.83        &0.68        &-0.17    &  -0.12         &-0.36$\pm$    0.11 \\
&1.67        &1.60        &-0.15    &  0.04         &-0.83$\pm$    0.15 \\
&1.67        &1.60        &-0.5     &  0.04        &-0.67$\pm$    0.13 \\
&1.20       &1.00        &0.21      &  0.20       &-0.26$\pm$    0.11 \\

Error  &1.304	 &1.202	 & 0.180 &0.112 & 0.436 \\ \hline

H$_2$O  
&1.04          &0.98         &-0.10        & -0.07       &-0.19 $\pm$ 0.24 \\
&(1.07)         &(1.13)         &(2.21)   &    (2.24)           &(-0.07) \\
&0.76        & 0.69        &-0.05   & -0.07            &0.03 $\pm$ 0.24 \\
&1.01       & 0.94        &-0.10    & -0.08          &-0.22 $\pm$ 0.26 \\
&0.70       & 0.61        &-0.06    & -0.08           &-0.073 $\pm$ 0.26 \\
&0.20       & 0.15        &-0.05    & -0.03          &-0.63 $\pm$ 0.23 \\
&0.42       & 0.20        &-0.10    & -0.08          &-0.59 $\pm$ 0.23 \\
&1.43       & 1.40        &0.002    & 0.14          &-0.33 $\pm$ 0.34 \\
&1.54       & 1.46        &0.14     & 0.18         &-0.33 $\pm$ 0.30 \\

Error  &     0.888	& 0.804	& 0.074	 &0.093 & 0.308 &\\ \hline  \hline

\end{tabular} 
\end{center}
\caption{Difference of the lowest singlet energy (in eV) from CIS, TDHF, CIS(D), EOM-CC2, and GF2-BSE  against EOM-CCSD for a list of molecules. In parenthesis, we show the difference of binding energies calculated for the lowest excitation for each method against EOM-CCSD. }
\label{table:four}
\end{table}

We first examine the vertical excitations LiH, HF, and $\mbox{H}_2\mbox{O}$, as shown in Table~\ref{table:four}. We find that for the entire range of excitations, the GF2-BSE approach outperforms CIS and TDHF, with an overall error that is smaller by a factor of $3$ compared to the other two methods. Comparing the results obtained by the GF2-BSE with the CIS(D) and EOM-CC2, we find that for the lower excitation energies, the two approach provide similar accuracies, while the CIS(D) and EOM-CC2 provide a better description of higher vertical excitations. This seems to follow the trends found for atoms and for $\mbox{H}_2$ molecules, however, unlike the case of the smaller system, where only the lowest singlet excitation was accurately described by the GF2-BSE approach, in the case of larger molecules, the GF2-BSE approach provides an accurate description for several low singlet excitations. 

Next, we examine the vertical excitations in two other molecules, $\mbox{CH}_4$ and LiF, as shown in Table~\ref{table:3}. Regarding the former molecule, we find the GF2-BSE under-performs even in comparison to CIS and TDHF. However, comparing the binding energies calculated for the lowest excitation (shown in parenthesis for each method), we find that GF2-BSE agrees well with the EOM-CCSD approach, while CIS/TDFH deviate significantly. This suggests that the main source of error in the GF2-BSE approach is the estimation of the fundamental gap and indeed, we find that the GF2 underestimates the ionization potential by more than an eV for $\mbox{CH}_4$.\cite{dou2019stochastic} The more accurate excitation energies obtained from CIS/TDHF in $\mbox{CH}_4$ seems to result from a cancellation of errors in the estimation of the fundamental gaps (by ~$1.2$~eV) and the excitonic binding energy (by ~$1$~eV). 

\begin{table} 
\caption{(Continue) Difference of the lowest singlet energy (in eV) from CIS, TDHF, CIS(D), EOM-CC2, and GF2-BSE  against EOM-CCSD for the listed molecules. In parenthesis, we show the difference of binding energies calculated for the lowest excitation for each method against EOM-CCSD.}
\begin{tabular}{ c c c c c c c c} 
 \hline \hline
           &   $\Delta$CIS      &$\Delta$TDHF      &  $\Delta$CIS(D)  &  $\Delta$EOM-CC2  & $\Delta$GF2-BSE & \\ \hline 
                               
CH$_4$  
&0.44     & 0.42       & 0.12   &0.08  &  -1.06 $\pm$      0.18 &\\
&(0.80)         &(0.82)  &(1.11)      & (1.16)          &(-0.19) \\
&0.44     & 0.42       & 0.12  &0.08   & -0.94 $\pm$      0.17 &\\
&0.44     & 0.42     & 0.12     &0.08  & -0.80$\pm$      0.17 &\\
&0.54     & 0.51      & 0.15   &0.14       & -0.87$\pm$      0.18 &\\
&0.53     & 0.51       & 0.15  &0.14     & -0.80 $\pm$      0.18 &\\
&0.53     & 0.51       & 0.15  &0.14   & -0.72 $\pm$      0.18 &\\
&0.52     & 0.49       & 0.12  &0.11     &-0.93$\pm$      0.18 &\\
&0.52     & 0.49       & 0.12  &0.11     & -0.82$\pm$      0.17 &\\
&0.44     & 0.37       & 0.04  &0.07     & -0.82 $\pm$      0.17 &\\ 
Error  &      0.490	& 0.459	& 0.122	&0.10  & 0.868 &\\ \hline
             
LiF  
&1.82        &1.82         &-1.42  &-0.93 & -0.75$\pm$     0.09 \\
&(-0.02)     &(-0.01)      &(3.23) & (2.74) & (0.15)               \\
&1.82         &1.82         &-1.42  &-0.93 &-0.62 $\pm$     0.08 \\
&2.05          &2.04         &-1.63 &-1.12 &-0.93 $\pm$     0.10 \\
&1.56         &1.52         &-1.07 &-0.78 &-0.46 $\pm$     0.08 \\
&1.58        &1.55          &-1.15 &-0.78  &-0.39 $\pm$     0.08 \\
&1.54         &1.52          &-1.18 &-0.78  &-0.36 $\pm$     0.08 \\
&1.78         &1.75         &-1.32 &-1.03 &-0.65 $\pm$     0.10 \\
&1.68         &1.65         &-1.43 &-1.06 &-0.71$\pm$     0.10 \\
&1.66        &1.60        &-1.26 &-0.81 &-0.52$\pm$     0.08 \\

Error  &      1.723 &	1.698 &	1.319 &0.914 & 0.600 \\ \hline\hline
\end{tabular} \label{table:3}
\end{table}

While the overall comparison of all methods with EOM-CCSD for LiF yields rather poor results compared  to the case of LiH, HF, and \ce{H2O}, the GF2-BSE approach seems to systematically be more accurate than CIS, TDHF, EOM-CC2, and CIS(D) for all excitations energies. The source of deviation from the EOM-CCSD approach for all methods can be traced to errors in estimating the fundamental gaps (all binding energies are very similar), where the GF2 approach provides a more accurate description leading to better agreement for the excitations energies. 

\begin{figure}[t] 
   \centering{}
      \includegraphics[width=10cm]{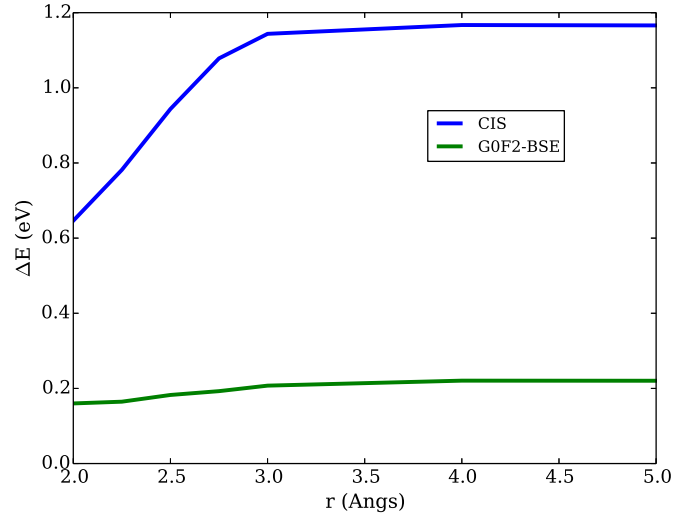} 
   \caption{Error of S4 for He dimer using G0F2-BSE, and CIS as compared to EOM-CCSD.}
   \label{fig:2}
\end{figure}

To better illustrate these conclusions, we plot in Fig.~\ref{fig:2} the excited state energies listed in Table \ref{table:3} and \ref{table:four} versus the EOM-CCSD results. Note that CIS(D) and GF2-BSE results are in much better agreement with the EOM-CCSD results compared to CIS/TDHF (the latter is not shown, but provides similar results to CIS). Overall, GF2-BSE method performs much better than CIS/TDHF and for the lower excited state energies it provides similar results in comparison to CIS(D).

\begin{figure}[t] 
   \begin{center}
   \includegraphics[width=10cm]{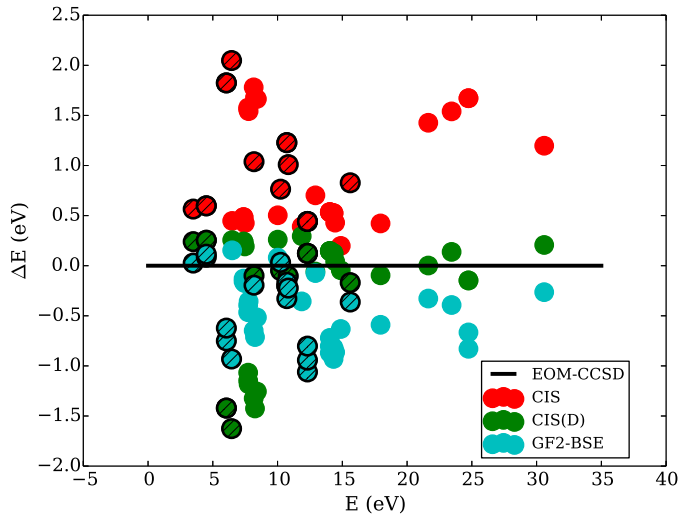} 
   \caption{Excitation energies of a few lowest states for a list of molecules listed in Table \ref{table:3} and \ref{table:four}. We use EOM-CCSD results as our reference (black line). Note that GF2-BSE results are in good agreement with CCSD results, better than CIS and are comparable with CIS(D) results. {\color{black} EOM-CC2 results (not shown) are very similar to CIS(D) results. The circles with slashes denote the lowest 3 singlet energies of the listed molecules.} }
   \end{center}
   \label{fig:1}     
\end{figure}

To further analyze the GF2-BSE approach, we have also calculated the lowest singlet excitation energy for He dimer. For the lowest two singlets excitations, we find that the excitation energies are independent of the distance between the two He atoms for large separation, consistent with previous calculations.\cite{subotnik2011communication} They are governed by local excitation of each He atom.  The third and forth excitations are characterized by charge transfer between the two He atoms and thus, show significant depends on the separation between the two He atoms, even at large distances. In Fig.~\ref{fig:2} we show that for the charge transfer state (state number 4), CIS predicts the overall correct trend, but the magnitude of the excitation energies can be off by an eV compared to the EOM-CCSD. Indeed, in Ref.~\citenum{subotnik2011communication}, Subotnik points out that CIS (or TDHF) has a large bias suppressing charge transfer character. Our G0F2-BSE method provides results for the transfer state that is in very good agreement with EOM-CCSD results. 

\section{Conclusions} \label{sec:con}
We have developed a time dependent second order Green's function theory to describe the response of a many-body molecular system to an external driving force. In principle, the framework can be combined with the stochastic resolution of identity to reduce the scaling of to $O(N^3)$ in both the weak and strong driving fields. In the present work, we have further introduced a linear-response approach for weak driving fields, which allowed us to recast the equation of motion for the time dependent second order Green's function into a frequency domain Casida-like equation (GF2-BSE). We find that the approach provides more accurate vertical excitation energies compared to the CIS and TDHF and is competitive with CIS(D) for a broad class of molecular systems, particularly for charge transfer excitations. Moreover, the GF2-BSE approach provides more accurate binding energies compared to the other methods, suggesting that the main source of error in the GF2-BSE approach results from under/over estimation of the fundamental gaps, implying that the approach provides a reasonably accurate framework to describe electron-hole correlation in molecular systems. Given the facile route to construct a low-scaling sGF2 approach to neutral excitations, we believe GF2-BSE should be the current method of choice for binding energies in molecules.  In addition, with corrected fundamental gaps, the approach should also be competitive with the most sophisticated approaches for calculating the location of low-lying excited states as well. Future work will be devoted to a full sGF2-BSE implementation as well as the reporting of a complete benchmark set of results for a wide range of molecular systems.

\acknowledgements
We would like to thank Yang-Hao Chen, Daniel Neuhauser, Vojtech Vlcek for helpful discussions. RB  gratefully acknowledges support from the US-Israel Binational Science Foundation BSF-201836. ER and DRR are grateful for support by the U.S. Department of Energy, Office of Science, Office of Advanced Scientific Computing Research, Scientific Discovery through Advanced Computing (SciDAC) program, under Award No. DE-SC0022088. Resources of the National Energy Research Scientific Computing Center (NERSC), a U.S. Department of Energy Office of Science User Facility operated under Contract No. DE-AC02-05CH11231 are greatly acknowledged. 

\appendix

\section{Evaluation of $\tilde{\Sigma}^{\rm ad}(t)$ and $\delta \tilde W^R_{ikjl} (\omega)$} \label{app:a}
In this Appendix, we provide more details concerning the evaluation of $\tilde{\Sigma}^{\rm ad}(t)$ and $\delta \tilde W^R_{ikjl} (\omega)$ appearing Sec. \ref{sec:theory} and Sec. \ref{sec:weak} of the main text. The retarded self-energy in the second order Born approximation is given by Eq. \ref{eq:sigmaR}. We have also defined $\delta W^R_{ikjl}(t_1, t_2 )$ and $\delta W^>_{ikjl}(t_1, t_2 )$ in Eq. \ref{eq:WR}. The retarded self-energy given in Eq.~\eqref{eq:sigmaR} has two terms when Langreth rules are employed. The first term can be evaluated at $t=0$ to give
\begin{eqnarray}
&&\tilde \Sigma^{\rm ad1}_{ij}( t=0, \omega ) =  i \sum_{kl} \int \frac{d\omega_1 }{2\pi} \delta \tilde W^R_{ikjl}(\omega - \omega_1) G_{kl}^<(\omega_1)  \nonumber \\ 
&&=  i \sum_{kl} \int \frac{d\omega_1}{2\pi}  \delta \tilde W^R_{ikjl}( \omega - \omega_1) 2\pi i \delta(\omega_1 - E_k) f(\omega) \delta_{kl} \nonumber \\
&&=  i \sum_{kl} \delta \tilde W^R_{ikjl}(\omega - E_k) i f(E_k) \delta_{kl}.
 \end{eqnarray}
 To obtain $\tilde \Sigma^{\rm ad1}_{ij}( t, \omega )$ at any time, we replace $i f(E_k) \delta_{kl}$ by $G^<(t, \tau=0)$, such that
\begin{eqnarray}
\tilde \Sigma^{\rm ad1}_{ij}( t, \omega ) = i \sum_{kl} \delta \tilde W^R_{ikjl}(\omega - E_k) G^<_{kl} (t, \tau=0).
 \end{eqnarray}
Similar steps can be taken for the second term in Eq.~\eqref{eq:sigmaR}, 
 \begin{eqnarray} 
&&\tilde \Sigma^{\rm ad2}  (t=0, \omega)  =  i \sum_{kl} \int \frac{d\omega_1}{2\pi}  \tilde G^R_{kl}(\omega- \omega_1 ) \delta \tilde W^>_{ikjl} ( \omega_1 ) \nonumber \\
=&& i \sum_{kl} \int_{-\infty}^\infty  \frac{d\omega_1}{2\pi}  \frac{1}{\omega - \omega_1 - E_k + i\eta}  \delta_{kl}   2i {\rm Im}  \delta \tilde W^R_{ikjl} (\omega_1) \Theta(\omega_1). \nonumber \\
\end{eqnarray} 
In the above, we have used the known relation for the non-interacting GF, $\tilde G^<_{mn}(t=0, \omega )  = i 2\pi \delta( \omega -E_n) f(\omega) \delta_{mn} $, as well as the fluctuation-dissipation theorem at zero temperature
 \begin{eqnarray}
 \delta \tilde W^>_{ikjl} ( \omega_1 )  =  2i {\rm Im} \delta \tilde W^R_{ikjl} (\omega_1)  \Theta (\omega_1),
 \end{eqnarray} 
where $\Theta (\omega_1)$ is a step function. Taking the real part of the above equation and using the Kramers-Kronig relationship, we obtain
 \begin{eqnarray} 
\tilde \Sigma^{\rm ad2}_{ij}  (t=0, \omega) &&=  \Re \sum_{kl} \delta \tilde W^R_{ikjl} (\omega-E_k)   \Theta(\omega-E_k)   \delta_{kl}  \nonumber \\
&&\approx  \frac12  \Re \sum_{kl} \delta \tilde W^R_{ikjl} (0) \delta_{kl}. 
\end{eqnarray} 
Similarly, we can evaluate $\delta \tilde W^R_{ik,jl} (\omega)$ appearing in the above equations.  We first Fourier transform Eq.~\eqref{eq:WR} to obtain
 \begin{eqnarray}
&&\delta \tilde W^R_{ikjl} (\omega) = -i \sum_{mnpq} \int \frac{d\omega_1}{2\pi} \Big( \tilde G^<_{mn}( \omega ) { \tilde G^A_{qp}} (  \omega - \omega_1)   \nonumber \\
&&+ \tilde G^R_{mn}( \omega ) {\tilde G^<_{qp} } ( \omega -\omega_1 )  \Big)v_{impk} (2 v_{jnql} - v_{jlqn} ).
\end{eqnarray}
Using  the non-interacting Green's functions, we arrive at the working expression for $\delta \tilde W^R_{ikjl} (\omega)$ in Eq.\ref{eq:W}.

\end{document}